\documentclass[journal]{IEEEtran}

\usepackage{amsfonts}
\usepackage{amsmath}
\usepackage{amssymb}
\usepackage{cite}
\usepackage{graphicx}
\usepackage[bookmarks=false]{hyperref}
\usepackage[inline]{enumitem}
\usepackage{multirow}
\usepackage{url}
\usepackage{textcomp}
\usepackage{wasysym}
\graphicspath{figs}

\begin{document}
    \bstctlcite{BSTcontrol}
    \setlength{\abovedisplayskip}{3pt}
    \setlength{\belowdisplayskip}{3pt}

    \title{Text-based Audio Retrieval by Learning from Similarities between Audio Captions}

    \author{Huang Xie, Khazar Khorrami, Okko R\"as\"anen, Tuomas Virtanen
    \thanks{Manuscript created May 25, 2024; revised September 25, 2024.}% <-this % stops a space
    \thanks{The authors are with the Faculty of Information Technology and Communication Sciences, Tampere University, Tampere, 33720, Finland (e-mail: huang.xie@tuni.fi; khazar.khorrami@tuni.fi; okko.rasanen@tuni.fi; tuomas.virtanen@tuni.fi).}}

    \markboth{Journal of \LaTeX\ Class Files, Vol. 14, No. 8, August 2015}{Shell \MakeLowercase{\textit{et al.}}: Bare Demo of IEEEtran.cls for IEEE Journals}
    \maketitle

    \begin{abstract}
        This paper proposes to use similarities of audio captions for estimating audio-caption relevances to be used for training text-based audio retrieval systems.
        Current audio-caption datasets (e.g., Clotho) contain audio samples paired with annotated captions, but lack relevance information about audio samples and captions beyond the annotated ones.
        Besides, mainstream approaches (e.g., CLAP) usually treat the annotated pairs as positives and consider all other audio-caption combinations as negatives, assuming a binary relevance between audio samples and captions.
        To infer the relevance between audio samples and arbitrary captions, we propose a method that computes non-binary audio-caption relevance scores based on the textual similarities of audio captions.
        We measure textual similarities of audio captions by calculating the cosine similarity of their Sentence-BERT embeddings and then transform these similarities into audio-caption relevance scores using a logistic function, thereby linking audio samples through their annotated captions to all other captions in the dataset.
        To integrate the computed relevances into training, we employ a listwise ranking objective, where relevance scores are converted into probabilities of ranking audio samples for a given textual query.
        We show the effectiveness of the proposed method by demonstrating improvements in text-based audio retrieval compared to methods that use binary audio-caption relevances for training.
    \end{abstract}


    \IEEEpeerreviewmaketitle

    \section{Introduction}\label{sec:introduction}

    \IEEEPARstart{T}{ext-based} audio retrieval, aiming at retrieving audio data based on textual queries, has drawn increasing attention in recent years~\cite{Xie2022Language, Primus2022Improving, Weck2022Matching}.
    It has great potential in real-world applications, e.g., search engines and multimedia databases.
    Recent works have focused on contrastive learning approaches (e.g., CLAP~\cite{Wu2023Large}) utilizing large audio-caption datasets (e.g., WavCaps~\cite{Mei2023WavCaps}, Auto-ACD~\cite{Sun2023ALarge}).
    For instance,~\mbox{Primus~\textit{et al.}~\cite{Primus2023Advancing}} built their system upon large-scale contrastive learning with audio-caption pairs, resulting in enhanced performance in DCASE 2023 Challenge~\cite{dcase2023challenge}.

    Contrastive learning~\cite{Wu2023Large, Primus2023Advancing} operates with positive and negative audio-caption pairs, assuming a binary relevance between audio samples and captions.
    Audio-caption pairs are considered positive if the caption accurately describes the paired audio sample; otherwise, they are deemed negative.
    Those approaches aim to learn audio and caption representations in a shared embedding space, which will allow measuring audio-caption relevances to be used in retrieval.
    Likewise, current audio-caption datasets~\cite{Mei2023WavCaps, Sun2023ALarge, Kim2019AudioCaps, Drossos2020Clotho} comprise pairs of audio samples and their annotated captions, i.e., only positive pairs.
    To obtain large quantities of negative pairs for contrastive learning, all other audio-caption combinations in the dataset are utilized as negative pairs.

    \begin{table}[!t]
        \caption{Caption Examples from AudioCaps}
        \label{tab:caption_examples}
        \setlength{\tabcolsep}{3pt}
        \renewcommand{\arraystretch}{1.2}
        \centering
        \begin{tabular}{l|c}
            \hline
            Audio Caption                                       & Textual Similarity \\
            \hline
            A cat is crying and a person is speaking            & 1.00               \\
            \hline
            A man is talking and a cat crying                   & 0.90               \\
            \hline
            A cat is crying                                     & 0.80               \\
            \hline
            A woman speaks then a cat sighs                     & 0.78               \\
            \hline
            People speak to each other, and a cat wails angrily & 0.77               \\
            \hline
            \multicolumn{2}{p{250pt}}{Each caption describes an individual audio sample in AudioCaps~\cite{Kim2019AudioCaps}. The caption ``A cat is crying and a person is speaking'' is used as the reference for similarity calculation. Textual similarity is measured with the cosine similarity between Sentence-BERT~\cite{Reimers2019Sentence} embeddings of captions.} \\
        \end{tabular}
        \vspace{-12pt}
    \end{table}

    It is likely that audio-caption datasets contain semantically similar captions for different audio samples.
    For instance, Table~\ref{tab:caption_examples} showcases multiple captions from AudioCaps~\cite{Kim2019AudioCaps}, each describing an audio sample that contains cat meows and (or) human speech.
    These captions exhibit notable textual similarities, e.g., high cosine similarities between their Sentence-BERT~\cite{Reimers2019Sentence} embeddings.
    Constructing negative audio-caption pairs from these captions and their respective audio samples may lead to false negatives, thereby potentially hindering the performance of trained systems.
    Therefore, instead of binary relevances, employing non-binary measures (e.g., graded relevance~\cite{Sakai2021Graded, Roitero2021On}) or to model partial relevance is essential for accurately portraying the relationship between audio samples and captions.

    On the other hand, limited research has been conducted on assessing the relevance between audio samples and captions in current audio-caption datasets.
    Our previous works~\cite{Xie2023Crowdsourcing, Xie2024Integrating} graded audio-caption relevances for a limited subset of Clotho~\cite{Drossos2020Clotho} via human crowdsourced assessments, which are usually labor-intensive.
    Recent work~\cite{Primus2024Estimated} estimated audio-caption relevances by aggregating predictions from multiple audio-language models trained with contrastive learning (i.e., positive and negative pairs), at the cost of increased computational complexity.

    \begin{figure*}[!t]
        \centering
        \includegraphics[width=1.0\textwidth]{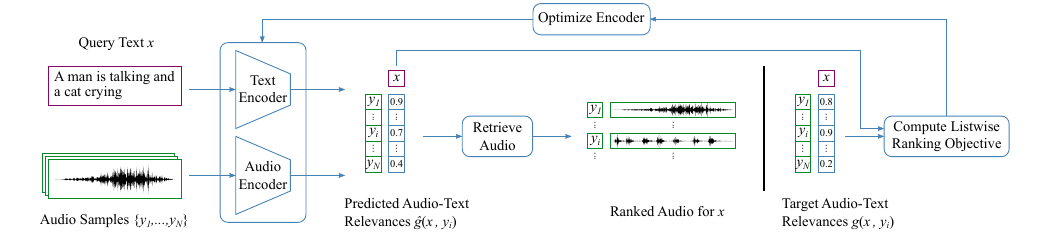}
        \caption{A model-agnostic dual-encoder framework for text-based audio retrieval. The query text $x$ and audio samples $y_{i} \in \left\{ y_{1},\cdots,y_{N} \right\}$ are projected into a shared embedding space by utilizing a dual-encoder model, and the cosine similarity of their embeddings is utilized as a prediction of their relevance, denoted as $\hat{g}(x,y_{i})$. Audio retrieval is performed by ranking $y_{1},\cdots,y_{N}$ by their predicted relevance to $x$. The dual-encoder model is trained by optimizing a listwise ranking objective, which is computed based on the predicted relevances $\hat{g}(x,y_{i})$ and the target relevances $g(x,y_{i})$.}
        \label{fig:audio_retrieval_framework}
        \vspace{-9pt}
    \end{figure*}

    In this work, we propose a method for computing non-binary audio-caption relevances within audio-caption datasets based on the textual similarities of audio captions, and then utilize the computed relevances to train models for text-based audio retrieval.
    Specifically, we measure textual similarities of audio captions by calculating the cosine similarity of their Sentence-BERT~\cite{Reimers2019Sentence} embeddings, and then transform textual similarities into audio-caption relevance scores using a logistic function.
    Subsequently, we utilize a listwise ranking objective (i.e., ListNet~\cite{Cao2007Learning}) to integrate the computed relevances into training.
    We show the effectiveness of the proposed method by demonstrating improvements in text-based audio retrieval compared to methods that use binary relevances for training.
    The proposed method can be applied to any captioning dataset (e.g., computing relevances between images and captions in image-caption datasets for visual-textual learning~\cite{Li2022Image, Li2024Integrating}) and, in addition to retrieval, could also be used in other tasks that are based on cross-modal similarity learning.

    The remainder of this paper is organized as follows.
    Section~\ref{sec:text-based-audio-retrieval} formalizes the problem of text-based audio retrieval.
    The proposed method is then presented in Section~\ref{sec:proposed-method}, followed by experimental setup in Section~\ref{sec:experiments} and corresponding results in Section~\ref{sec:results-and-analysis}.
    Finally, we conclude this paper in Section~\ref{sec:conclusions}.

    \section{Text-based Audio Retrieval}\label{sec:text-based-audio-retrieval}

    Text-based audio retrieval refers to retrieving relevant audio samples from a dataset given a textual query.
    In practice, solving the problem is usually done by estimating a relevance score for each sample to the query and ranking them by relevance.
    We formalize text-based audio retrieval as follows.
    Given a query text $x$ and a set of $N$ audio samples $Y=\left\{ y_{1},\cdots,y_{N} \right\}$, the task is to arrange $y_{i} \in Y$ in descending order of their relevance to $x$.
    We define the relevance of $y_{i}$ to $x$ as $g(x,y_{i})$, and our aim is to develop a model that can accurately predict $g(x,y_{1}),\cdots,g(x,y_{N})$.

    As illustrated in Fig.~\ref{fig:audio_retrieval_framework}, we employ a model-agnostic dual-encoder framework.
    A dual-encoder model is trained to learn representations of $x$ and $y_{i}$ in a shared embedding space, where cosine similarity is calculated as the prediction of $g(x,y_{i})$, denoted as $\hat{g}(x,y_{i})$.
    During training, we optimize a listwise ranking objective, which is computed between the target relevances $g(x,y_{i})$ and their predictions $\hat{g}(x,y_{i})$, for $i=1,\cdots,N$.
    For audio retrieval, we rank $y_{1},\cdots,y_{N}$ by their predicted relevance to $x$.

    Similar to previous studies~\cite{Wu2023Large, Primus2023Advancing}, we utilize audio-caption datasets for training.
    Suppose that the training dataset is $D=\left\{ (x_{1},y_{1}),\cdots,(x_{N},y_{N}) \right\}$, where $x_{i}$ is the caption of $y_{i}$, for $i=1,\cdots,N$.
    During training, each caption $x_{i}$ is used as a query, and the dual-encoder model is trained to produce the target relevances $g(x_{i},y_{j})$, for $j=1,\cdots,N$.

    Previous studies~\cite{Wu2023Large, Primus2023Advancing} assumed a binary relevance between $x_{i}$ and $y_{j}$: $g(x_{i},y_{j})=1$ if $i=j$ and otherwise $g(x_{i},y_{j})=0$.
    However, the training dataset can include multiple audio samples matching with a single caption (and vice versa), i.e., $g(x_{i},y_{j})=1$ for some $i \neq j$.
    Besides, there can be varying levels of relevance between $x_{i}$ and $y_{j}$, ranging from fully relevant, e.g., $g(x_{i},y_{j})=1$, to partially relevant, e.g., $g(x_{i},y_{j})=0.5$.
    Therefore, using $g(x_{i},y_{j}) \in \left\{ 0,1 \right\}$ for training may result in a suboptimal solution.
    In this work, we compute $g(x_{i},y_{j}) \in \left[ 0,1 \right]$ based on the textual similarity of $x_{i}$ and $x_{j}$, and then integrate it into training by using a listwise ranking objective (Section~\ref{subsec:listwise-ranking-objective}).

    \section{Proposed Method}\label{sec:proposed-method}

    In this section, we first introduce the computation of audio-caption relevances based on the textual similarities of audio captions, and then present the listwise ranking objective, which integrates the computed relevances into training.

    \subsection{Computed Audio-Caption Relevance}\label{subsec:computed-audio-caption-relevance}

    Current audio-caption datasets (e.g., Clotho~\cite{Drossos2020Clotho}) consist of audio samples paired with annotated captions, but lack relevance information about audio samples and captions beyond the annotated ones.
    To infer this relevance information, we leverage the textual similarity of audio captions.

    In practice, audio captions exhibiting high mutual textual similarity can be appropriate descriptions of each other's audio contents.
    By analyzing the textual similarity between captions, we can determine potential agreement about audio content across audio-caption pairs.
    This enables us to identify implicit relevance between audio samples and captions from different pairs.
    For instance, consider the audio-caption pairs $(x_{i},y_{i})$ and $(x_{j},y_{j})$ with $i \neq j$.
    High textual similarity of $x_{i}$ and $x_{j}$ would then suggest that both pairs contain similar audio content, implying relevance between $x_{i}$ and $y_{j}$ (as well as between $x_{j}$ and $y_{i}$).
    Therefore, we explore computing $g(x_{i},y_{j})$ based on the textual similarity of $x_{i}$ and $x_{j}$.

    With the success of Sentence-BERT in semantic textual similarity tasks~\cite{Reimers2019Sentence}, we utilize it to gauge the textual similarity between audio captions.
    Specifically, we compute 768-dimensional caption embeddings and calculate their cosine similarity as a measure of textual similarity.
    We denote the textual similarity of $x_{i}$ and $x_{j}$ as $h(x_{i},x_{j})$.

    To score the relevance of $y_{j}$ to $x_{i}$, we define $g(x_{i},y_{j})$ as a function of $h(x_{i},x_{j})$, written as
    \begin{equation}
        \label{eq:relevance_function}
        g(x_{i},y_{j}) = f(h(x_{i},x_{j})),
    \end{equation}
    where $f$ is a monotonically non-decreasing function over the interval $\left[ -1,1 \right]$.
    In practice, we experimented with linear and logistic functions for $f$ (Section~\ref{subsec:the-selection-of-function-f}).

    \subsection{Listwise Ranking Objective}\label{subsec:listwise-ranking-objective}

    To integrate the computed relevances into training, we employ a listwise ranking objective, i.e., ListNet~\cite{Cao2007Learning}.
    Given the computed relevances $G_{i}=\left\{ g(x_{i},y_{j}) \arrowvert y_{j} \in Y \right\}$ and the model-predicted relevances $\hat{G}_{i}=\left\{ \hat{g}(x_{i},y_{j}) \arrowvert y_{j} \in Y \right\}$ for $x_{i} \in D$, we calculate two probability distributions over $Y$ and then compute their cross-entropy as the ListNet loss.

    For $G_{i}$, we define a probability distribution $P$, written as
    \begin{equation}
        \label{eq:probability_P}
        p(y_{j} \arrowvert x_{i}) = \dfrac{e^{g(x_{i},y_{j}) / \omega}}{\sum_{k=1}^{N} e^{g(x_{i},y_{k})  / \omega}},
    \end{equation}
    where $\omega$ is a temperature parameter and $y_{j} \in Y$, for $j=1,\cdots,N$.
    The $p(y_{j} \arrowvert x_{i})$ is interpreted as the probability of ranking $y_{j}$ highest among $Y$ for $x_{i}$~\cite{Xie2024Integrating, Cao2007Learning}.
    Similarly, we compute a probability distribution $Q$ from $\hat{G}_{i}$, written as
    \begin{equation}
        \label{eq:probability_Q}
        q(y_{j} \arrowvert x_{i}) = \dfrac{e^{\hat{g}(x_{i},y_{j}) / \tau}}{\sum_{k=1}^{N} e^{\hat{g}(x_{i},y_{k})  / \tau}},
    \end{equation}
    where $\tau$ is a temperature parameter for $\hat{G}_{i}$.
    The ListNet loss of $G_{i}$ and $\hat{G}_{i}$ is calculated as
    \begin{equation}
        \label{eq:listnet_loss}
        L(P,Q) = - \sum_{j=1}^{N} p(y_{j} \arrowvert x_{i}) \log q(y_{j} \arrowvert x_{i}).
    \end{equation}

    At the training stage,~\eqref{eq:listnet_loss} is minimized to ensure that the model can accurately predict $G_{i}$ for every $x_{i} \in D$.
    During audio retrieval, the trained model produces relevance scores for audio samples with respect to a given query text.

    \begin{table*}[!t]
        \caption{Retrieval Performance on AudioCaps and Clotho. Mean and SD ($\pm$) are reported across five independent runs.}
        \label{tab:baseline_results}
        \setlength{\tabcolsep}{3pt}
        \renewcommand{\arraystretch}{1.2}
        \centering
        \begin{tabular}{l|l|cccc|cccc}
            \hline
            \multirow{2}{*}{Dataset}    & \multirow{2}{*}{Method}               & \multicolumn{4}{c|}{Text-Based Audio Retrieval}                                                               & \multicolumn{4}{c}{Audio-Based Text Retrieval}                                                                \\
            \cline{3-10}
                                        &                                       & mAP@10                    & R@1                       & R@5                       & R@10                      & mAP@10                    & R@1                       & R@5                       & R@10                      \\
            \hline
            \multirow{4}{*}{AudioCaps}  & InfoNCE                               &           54.5$\pm$0.5    &           39.6$\pm$0.7    &           75.4$\pm$0.4    &           86.9$\pm$0.2    &           37.0$\pm$0.5    &           9.7$\pm$0.2     &           38.0$\pm$0.7    &           54.4$\pm$0.5    \\
                                        & ListNet\textsubscript{audio}          & \bfseries 55.0$\pm$0.3    & \bfseries 39.9$\pm$0.3    & \bfseries 76.1$\pm$0.5    & \bfseries 87.6$\pm$0.4    &           31.4$\pm$0.1    &           8.5$\pm$0.2     &           32.5$\pm$0.3    &           48.7$\pm$0.2    \\
                                        & ListNet\textsubscript{text}           &           48.7$\pm$0.3    &           33.7$\pm$0.5    &           69.5$\pm$0.2    &           82.8$\pm$0.4    & \bfseries 37.8$\pm$0.2    & \bfseries 10.0$\pm$0.2    & \bfseries 38.6$\pm$0.3    &           54.8$\pm$0.4    \\
                                        & ListNet\textsubscript{audio+text}     &           54.7$\pm$0.1    &           39.8$\pm$0.2    &           75.4$\pm$0.4    &           87.0$\pm$0.3    &           37.4$\pm$0.4    &           9.9$\pm$0.1     &           38.1$\pm$0.4    & \bfseries 55.0$\pm$0.5    \\
            \hline
            \multirow{4}{*}{Clotho}     & InfoNCE                               &           28.2$\pm$0.5    &           16.9$\pm$0.5    &           43.3$\pm$0.3    &           57.7$\pm$0.7    &           15.1$\pm$0.3    &           4.3$\pm$0.2     &           17.0$\pm$0.3    &           26.9$\pm$0.4    \\
                                        & ListNet\textsubscript{audio}          & \bfseries 30.4$\pm$0.1    & \bfseries 19.0$\pm$0.1    & \bfseries 45.9$\pm$0.3    & \bfseries 60.1$\pm$0.5    &           16.0$\pm$0.2    &           4.3$\pm$0.1     &           17.8$\pm$0.3    &           27.9$\pm$0.2    \\
                                        & ListNet\textsubscript{text}           &           28.5$\pm$0.4    &           17.3$\pm$0.4    &           43.6$\pm$0.3    &           58.3$\pm$0.2    & \bfseries 16.5$\pm$0.3    & \bfseries 4.4$\pm$0.2     & \bfseries 18.7$\pm$0.2    & \bfseries 28.8$\pm$0.4    \\
                                        & ListNet\textsubscript{audio+text}     &           29.6$\pm$0.1    &           18.0$\pm$0.1    &           45.3$\pm$0.2    &           59.4$\pm$0.3    &           16.0$\pm$0.2    & \bfseries 4.4$\pm$0.1     &           17.9$\pm$0.1    &           28.0$\pm$0.4    \\
            \hline
        \end{tabular}
        %\vspace{-9pt}
    \end{table*}
    \begin{table}[!t]
        \caption{Retrieval Performance on AudioCaps and Clotho with Pretraining and Fine-tuning}
        \label{tab:finetuned_results}
        \setlength{\tabcolsep}{3pt}
        \renewcommand{\arraystretch}{1.2}
        \centering
        \begin{tabular}{l|l|cccc}
            \hline
            \multirow{2}{*}{Dataset}    & \multirow{2}{*}{Method}               & \multicolumn{4}{c}{Text-Based Audio Retrieval}                                                                \\
            \cline{3-6}
                                        &                                       & mAP@10                    & R@1                       & R@5                       & R@10                      \\
            \hline
            \multirow{2}{*}{AudioCaps}  & InfoNCE                               &           55.5$\pm$0.2    &           40.8$\pm$0.3    &           76.0$\pm$0.3    &           87.4$\pm$0.2    \\
                                        & ListNet\textsubscript{audio}          & \bfseries 56.5$\pm$0.4    & \bfseries 41.9$\pm$0.5    & \bfseries 76.8$\pm$0.2    & \bfseries 87.9$\pm$0.4    \\
            \hline
            \multirow{2}{*}{Clotho}     & InfoNCE                               &           35.2$\pm$0.5    &           23.2$\pm$0.5    &           51.4$\pm$0.6    &           65.6$\pm$0.3    \\
                                        & ListNet\textsubscript{audio}          & \bfseries 36.2$\pm$0.3    & \bfseries 24.1$\pm$0.5    & \bfseries 52.6$\pm$0.3    & \bfseries 66.1$\pm$0.6    \\
            \hline
        \end{tabular}
        \vspace{-6pt}
    \end{table}

    \section{Experiments}\label{sec:experiments}

    The aim of the experiments was to test the proposed audio-caption relevance scoring in text-based audio retrieval.
    For this purpose, we used audio-caption datasets: AudioCaps~\cite{Kim2019AudioCaps}, Clotho~\cite{Drossos2020Clotho}, and WavCaps~\cite{Mei2023WavCaps}, with a dual-encoder model from~\cite{Primus2024Estimated}.

    \subsection{Audio-Caption Datasets}\label{subsec:audio-caption-datasets}

    AudioCaps~\cite{Kim2019AudioCaps} consists of 51,308 audio samples and 57,188 captions, split into three subsets: a training set with 49,838 audios, a validation set with 495 audios, and a testing set with 975 audios.
    All audios are drawn from YouTube videos, and their captions are crowdsourced from human annotators.
    One human-annotated caption is provided for each audio in the training set, and five captions are provided for each in the validation and test sets.

    We collected audios of AudioCaps from their original YouTube videos.
    Due to unavailable YouTube videos, we have 45,522 audios (91.3\%) for the training set, 449 audios (90.7\%) for the validation set, and 940 audios (96.4\%) for the test set.

    Clotho~\cite{Drossos2020Clotho} comprises 5,929 audio samples, each accompanied by five human-written captions, totaling 29,645 captions.
    All audios are sourced from the FreeSound~\cite{Font2013Freesound}, and their captions are crowdsourced using a three-step framework~\cite{Drossos2020Clotho}.
    Clotho is partitioned into three subsets: a development set with 3,839 audios, a validation set with 1,045 audios, and an evaluation set with 1,045 audios.

    WavCaps~\cite{Mei2023WavCaps} is a large-scale, weakly-labeled audio-caption dataset that contains over 400,000 audio samples, each accompanied by a GPT-generated caption.
    The audio samples are collected from FreeSound~\cite{Font2013Freesound}, BBC Sound Effects~\cite{BBCSoundEffects}, SoundBible~\cite{SoundBible}, and AudioSet Strongly-Labeled Subset~\cite{Hershey2021The}.
    We excluded the overlapping audio samples between WavCaps and Clotho, resulting in 401,195 audio-caption pairs.

    \subsection{Dual-Encoder Model}\label{subsec:dual-encoder-model}

    We used audio and text encoders from the best-ranked system~\cite{Primus2024Estimated} in DCASE 2024 Challenge~\cite{dcase2024challenge}.
    PaSST~\cite{Koutini2022Efficient} was used as the audio encoder, while RoBERTa large~\cite{Liu2019RoBERTa} served as the text encoder, each followed by two linear layers with a ReLU non-linearity in between, projecting audio samples and captions into a shared embedding space.
    The dual-encoder model was previously trained using binary audio-caption relevances with contrastive learning (e.g., InfoNCE~\cite{Oord2018Representation}).
    In contrast, this work trained it using computed non-binary relevances with listwise ranking.

    We followed the training stage one (without a model ensemble) from~\cite{Primus2024Estimated}.
    We first trained the dual-encoder model on AudioCaps and Clotho as baselines.
    We divided the training sets into mini-batches of 32 audio-caption pairs and trained the model with an Adam optimizer for 25 epochs.
    The learning rate was decayed from $2 \times 10^{-5}$ to $10^{-7}$ using cosine annealing~\cite{Loshchilov2017SGDR}.
    We experimented with large-scale pretraining by merging WavCaps with the training sets of AudioCaps and Clotho as a large training set and then fine-tuned the model on AudioCaps and Clotho.
    For pretraining and fine-tuning, we used the same configuration as for baselines.
    The temperature parameters $\omega$ and $\tau$ were set to $0.05$.

    \subsection{Audio-Based Text Retrieval}\label{subsec:audio-based-text-retrieval}

    Besides text-based audio retrieval, we adapted the listwise ranking objective (Section~\ref{subsec:listwise-ranking-objective}) to audio-based text retrieval (i.e., retrieving captions for audio samples) by swapping $x$ and $y$ in~\eqref{eq:probability_P},~\eqref{eq:probability_Q}, and~\eqref{eq:listnet_loss}.
    We trained the dual-encoder model for both retrieval scenarios.

    \subsection{Evaluation Metrics}\label{subsec:evaluation-metrics}

    Retrieval performance is evaluated in terms of mean Average Precision at 10 (mAP@10) and Recall at $k$ (R@$k$ with $k \in \left\{ 1,5,10 \right\}$), as done in~\cite{Primus2024Estimated}.
    The mAP@10 is calculated as the mean of Average Precision (AP) scores across all queries, with AP being the average of precisions at positions where relevant items appear in the ranked list of the top-10 retrieved items for a query.
    The R@$k$ is defined as the proportion of relevant items among the top-k items relative to the total relevant items of a query, which is then averaged over all queries.
    For both metrics, higher values indicate better performance.
    The evaluation is repeated five times, and the averaged metrics and their standard deviations are reported.

    \section{Results and Analysis}\label{sec:results-and-analysis}

    This section presents the experimental results of the proposed method on AudioCaps and Clotho.

    \subsection{The Selection of Function $f$}\label{subsec:the-selection-of-function-f}

    We experimented with the min-max scaling and logistic functions for $f$ in~\eqref{eq:relevance_function} and compared their results of text-based audio retrieval on Clotho.
    The logistic function was written as:
    \begin{equation}
        \label{eq:logistic_function}
        g(x_{i},y_{j}) = \frac{1}{1 + e^{2.73 - 4.58 \cdot h(x_{i},x_{j})}},
    \end{equation}
    which was derived by modeling the crowdsourced audio-caption relevance ratings~\cite{Xie2023Crowdsourcing} based on the similarities of audio captions through beta regression with a logit link function.
    It achieved superior performance ($30.4\pm0.1$ in mAP@10, $19.0\pm0.1$ / $45.9\pm0.3$ / $60.1\pm0.5$ in R@$\left\{ 1,5,10 \right\}$) compared to the min-max scaling ($29.0\pm0.6$ in mAP@10, $17.7\pm0.5$ / $44.4\pm0.6$ / $58.6\pm0.3$ in R@$\left\{ 1,5,10 \right\}$).
    Thus, we used the logistic function in the following experiments.

    \subsection{Audio and Text Retrieval}\label{subsec:audio-and-text-retrieval}

    Table~\ref{tab:baseline_results} summarizes retrieval performance on AudioCaps and Clotho.
    The model trained with computed non-binary relevances using~\eqref{eq:listnet_loss} is denoted as ``ListNet\textsubscript{audio}'', and the adapted version for audio-based text retrieval (Section~\ref{subsec:audio-based-text-retrieval}) is denoted as ``ListNet\textsubscript{text}''.
    We also experimented with the combination of the two, labeled ``ListNet\textsubscript{audio+text}''.
    The model trained with binary relevances using InfoNCE~\cite{Oord2018Representation} (the same as~\cite{Wu2023Large, Primus2023Advancing}) is denoted as ``InfoNCE''.

    The results demonstrated that the proposed method outperformed the InfoNCE approach in both retrieval scenarios on AudioCaps and Clotho.
    In particular, ListNet\textsubscript{audio} delivered the best performance in text-based audio retrieval, as it was specifically optimized for ranking audio samples based on text queries.
    Likewise, ListNet\textsubscript{text}, designed for ranking captions given an audio sample, achieved the highest performance in audio-based text retrieval.

    We assessed performance significance in text-based audio retrieval.
    A paired t-test was conducted on text queries by calculating mAP@10 for each across the five runs.
    The results showed that ListNet\textsubscript{audio} achieved significant improvements on Clotho ($t(5224)=11.658$, $p<0.001$) compared to InfoNCE.
    On AudioCaps, the improvement was not significant ($t(4699)=0.532$, $p>0.05$).
    %We hypothesize that this could be fixed with additional hyperparameter tuning for AudioCaps.

    Table~\ref{tab:baseline_results} shows that ListNet\textsubscript{audio+text} utilizing computed non-binary relevances outperforms \mbox{InfoNCE}, which uses binary relevances.
    We note that both losses can be viewed as the sum of two cross-entropy terms and share the same underlying formula when $P$ has all probability mass on one sample (Section~\ref{subsec:listwise-ranking-objective}).
    We conclude that calculating $P$ based on the non-binary relevances improves performance.
    %It indicates that leveraging audio-caption relevances derived from the similarities of audio captions improves performance.

    \subsection{Pretraining and Fine-tuning}\label{subsec:pretraining-and-fine-tuning}

    We also validated the effectiveness of the proposed method on large-scale data by pretraining the dual-encoder model on the three datasets, followed by fine-tuning on AudioCaps and Clotho, respectively.
    Table~\ref{tab:finetuned_results} summarizes the results of text-based audio retrieval on AudioCaps and Clotho.
%    Table~\ref{tab:finetuned_results} summarizes the results of text-based audio retrieval after pretraining the dual-encoder model on the three datasets, followed by fine-tuning on AudioCaps and Clotho, respectively.
    A paired t-test was conducted on text queries by calculating mAP@10 for each across the five runs.
    The results showed that ListNet\textsubscript{audio} achieved significant improvements in mAP@10 on both AudioCaps ($t(4699)=2.601$, $p=0.0093$) and Clotho ($t(5224)=2.438$, $p=0.0148$) compared to InfoNCE.

    \section{Conclusions}\label{sec:conclusions}

    This work proposed a method for computing audio-caption relevances based on the textual similarities of audio captions, and utilized the computed relevances to train models for text-based audio retrieval.
    We calculated relevance scores for audio samples and captions by transforming textual similarities using a logistic function.
    We employed a listwise ranking objective to integrate the computed relevances into training.
    Additionally, we experimented on audio-based text retrieval with the proposed method.
    Experimental results validated the effectiveness of the proposed method in both text-based audio retrieval and audio-based text retrieval, showcasing improvements over methods that use binary relevances for training.

    \ifCLASSOPTIONcaptionsoff
    \newpage
    \fi

    \bibliographystyle{IEEEtran}
    \bibliography{ms}

\end{document}